\documentclass[twocolumn,showpacs,aps,prb,superscriptaddress,floatfix,unsortedadresss]{revtex4-1}
\usepackage{graphicx}
\usepackage{dcolumn}
\usepackage{bm}
\usepackage[utf8x]{inputenc}
\usepackage[autostyle]{csquotes}
\usepackage{lineno}
\usepackage{xcolor}
\usepackage{soul}

\begin{document}

	\author{J.J. Wagman}
	\affiliation{Department of Physics and Astronomy, McMaster University, Hamilton, Ontario, Canada L8S 4M1}
	
	\author{J. P. Carlo}
	\affiliation{Department of Physics and Astronomy, McMaster University, Hamilton, Ontario, Canada L8S 4M1}
	\affiliation{Department of Physics, Villanova University, Villanova, PA 19085 USA}
	
	\author{J. Gaudet}
	\affiliation{Department of Physics and Astronomy, McMaster University, Hamilton, Ontario, Canada L8S 4M1}
	
	\author{G. Van Gastel}
	\affiliation{Department of Physics and Astronomy, McMaster University, Hamilton, Ontario, Canada L8S 4M1}
	
	\author{D. L. Abernathy}
	\affiliation{Quantum Condensed Matter Division, Oak Ridge National Laboratory, Oak Ridge, Tennessee 37831, USA}
	
	\author{M. B. Stone}
	\affiliation{Quantum Condensed Matter Division, Oak Ridge National Laboratory, Oak Ridge, Tennessee 37831, USA}
	
	\author{G.E. Granroth}
	\affiliation{Neutron Data Analysis and Visualization Division, Oak Ridge National Laboratory, Oak Ridge, Tennessee 37831, USA}
	
	\author{A. I. Koleshnikov}
	\affiliation{Chemical and Engineering Materials Division, Oak Ridge National Laboratory, Oak Ridge, Tennessee 37831, USA}
	
	\author{A. T. Savici}
	\affiliation{Neutron Data Analysis and Visualization Division, Oak Ridge National Laboratory, Oak Ridge, Tennessee 37831, USA}
	
	\author{Y. J. Kim}
	\affiliation{Department of Physics, University of Toronto, Toronto, Ontario, Canada M5S 1A7}
	
	\author{H. Zhang}
	\affiliation{Department of Physics, University of Toronto, Toronto, Ontario, Canada M5S 1A7}
	
	\author{D. Ellis}
	\affiliation{Department of Physics, University of Toronto, Toronto, Ontario, Canada M5S 1A7}
	
	\author{Y. Zhao}
	\affiliation{National Institute of Standards and Technology, Gaithersburg, Maryland 20899-6102, USA}
	\affiliation{Department of Materials Sciences and Engineering, University of Maryland, College Park, Maryland 20742, USA}
	
	\author{L. Clark}
	\affiliation{Department of Physics and Astronomy, McMaster University, Hamilton, Ontario, Canada L8S 4M1}
	
	\author{A.B. Kallin}
	\affiliation{Department of Physics and Astronomy, McMaster University, Hamilton, Ontario, Canada L8S 4M1}
	
	\author{E. Mazurek}
	\affiliation{Department of Physics and Astronomy, McMaster University, Hamilton, Ontario, Canada L8S 4M1}
	
	\author{H.A. Dabkowska}
	\affiliation{Brockhouse Institute for Materials Research, McMaster University, Hamilton, Ontario, Canada L8S 4M1}
	
	\author{B.D. Gaulin}
	\affiliation{Department of Physics and Astronomy, McMaster University, Hamilton, Ontario, Canada L8S 4M1}
	\affiliation{Brockhouse Institute for Materials Research, McMaster University, Hamilton, Ontario, Canada L8S 4M1}
	\affiliation{Canadian Institute for Advanced Research, 180 Dundas Street West, Toronto, Ontario, Canada M5G 1Z8}
	
	\begin{abstract}

We present time-of-flight neutron-scattering measurements on single crystals of $La_{2-x}Ba_{x}CuO_{4}$ (LBCO) with $0\leq $x$ \leq 0.095$ and $La_{2-x}Sr_{x}CuO_{4}$ (LSCO) with $x$ = 0.08 and 0.11. This range of dopings spans much of the phase diagram relevant to high temperature cuprate superconductivity, ranging from insulating, three dimensional (3D) commensurate long range antiferromagnetic order, for $x$ $\leq$ 0.02, to two dimensional (2D) incommensurate antiferromagnetism co-existing with superconductivity for $x$ $\geq$ 0.05. Previous work on lightly doped LBCO with $x$ = 0.035 showed a clear resonant enhancement of the inelastic scattering coincident with the low energy crossings of the highly dispersive spin excitations and quasi-2D optic phonons. The present work extends these measurements across the phase diagram and shows this enhancement to be a common feature to this family of layered quantum magnets. Furthermore we show that the low temperature, low energy magnetic spectral weight is substantially larger for samples with non-superconducting ground states relative to any of the samples with superconducting ground states. Spin gaps, suppression of low energy magnetic spectral weight as a function of decreasing temperature, are observed in both superconducting LBCO and LSCO samples, consistent with previous observations for superconducting LSCO.

\end{abstract}

\title{Neutron Scattering Studies of Spin-Phonon Hybridization and Superconducting Spin-Gaps in the High Temperature Superconductor $La_{2-x}(Sr,Ba)_{x}CuO_{4}$}

\maketitle

\section{Introduction}

There are several important similarities between different families of high temperature superconductors, which can also be common to certain low temperature superconductors\cite{Taillefer_AnnRevCondMattPhys_2010}. The most striking of these is the proximity of magnetism to superconducting ground states. Interestingly, the contiguous nature of these two ordered states has driven speculation that the two orders compete with each other, and also that magnetism may be intimately involved in the mechanism for Cooper pair formation in cuprate, iron-based, heavy fermion and organic superconductors\cite{Anderson_Science_1987,Emery_Nature_1995,Varma_Nature_2010,Mathur_Nature_1998,Sachdev_JPhysCondMat_2012,Savici_PRL_2005,Savici_PRB_2002}.

The 214 family of cuprate superconductors is the original family of high temperature superconductors to be discovered\cite{Bednorz_ZPhysB_1986}. Both $La_{2-x}Ba_{x}CuO_{4}$ (LBCO) and $La_{2-x}Sr_{x}CuO_{4}$ (LSCO) are relatively easy to grow as large and pristine single crystals, although the growth of the $La_{2-x}Sr_{x}CuO_{4}$ branch of the family is easier at higher $x$. As a result, this system has been extensively studied by techniques that require large single crystals, such as inelastic neutron scattering\cite{Wakimoto_PRL_2004}. However, advances in neutron scattering itself, and especially in time-of-flight neutron scattering at spallation neutron sources, have made it timely to revisit the spin and phonon dynamics in these systems, wherein sample rotation methods have allowed for the collection of comprehensive four dimensional data sets spanning {\bf Q} and $\hbar\omega$.

Both LBCO and LSCO lose their three dimensional commensurate (3D C) antiferromagnetic (AF) order quickly on doping with holes at finite $x$\cite{Wagman_PRB_2013,Keimer_PRB_1992}. This occurs at $x$ = 0.02 in both LSCO and LBCO. Quasi-two dimensional (2D) incommensurate short range frozen order replaces 3D C AF, with the onset of 2D order occuring at much lower temperatures, $\sim$ 25 K, for $x$ $\geq$ 0.02. As a function of increased doping, $x$, the wave-vector characterizing the 2D IC magnetism increases, consistent with the stripe picture introduced by Tranquada and co-workers\cite{Tranquada_Nature_1995}. Remarkably, the IC wave-vector rotates by 45 degrees, from so-called diagonal to parallel stripes at a doping level that is co-incident with the onset of a superconducting ground state, $x$ = 0.05 in both LBCO and LSCO\cite{Dunsiger_PRB_2008,Tranquada_AIP_2013}.

Independent of whether the AF order is C or IC, the quasi-2D spin excitations are known to be centered on two dimensional magnetic zone centers (2DMZCs), which are wave-vectors of the form ($\frac{1}{2}, \frac{1}{2}, L$), and equivalent wave-vectors. This notation implies a pseudotetragonal unit cell that is both convenient and appropriate given the relatively small orthorhombicity present in these materials\cite{Birgeneau_JPhysSocJpn_2006,Fujita_PRB_2002,Yamada_PRB_1998,Wagman_PRB_2015}. The quasi-2D spin excitations are also known to be highly dispersive and to extend to energies $\sim$ 200 - 300 meV depending on the precise level of doping\cite{Fujita_JPhysSocJpn_2012,Tranquada_JMagMM_2014,Coldea_PRL_2001,Headings_PRL_2010}. Recent time-of-flight neutron scattering on lightly doped, $x$ = 0.035, non-superconducting LBCO has revealed very interesting resonant enhancement of the magnetic spectral weight as a function of energy, that is co-incident with the low energy crossings of the highly dispersive spin excitations with weakly dispersive optic phonons\cite{Wagman_PRB_2015}. The optic phonon most strongly associated with this resonant enhancement, at $\sim$ 19 meV, could be identified with a breathing mode of (mostly) the oxygen ions within the CuO$_2$ planes. This phonon eigenvector is both quasi-2D itself, and is expected to couple strongly to the magnetism, as its displacements flex the main Cu-O-Cu superexchange pathway within the ab plane.

In this paper, we extend these and related time-of-flight neutron scattering measurements to other dopings in the LBCO and LSCO family, including several samples with sufficiently high doping to have superconducting ground states. These results show that the same phenomenology of resonant enhancement of the magnetic spectral weight at the low energy crossings of the very dispersive spin excitations with the weakly dispersive optic phonons, primarily at $\sim$ 15 and 19 meV, is a common feature across the phase diagram studied, from $x$ = 0 to $x$ = 0.11. We further show a common form for the energy dependence of $\chi\prime\prime({\bf Q},\hbar\omega)$ across this series at low temperatures, with non-superconducting samples showing greater weight at relatively low energies only, compared with samples with superconducting ground states. We also present evidence for a suppression of the low energy magnetic scattering within the superconducting ground state relative to the same scattering within the higher temperature normal state for both LBCO and LSCO. We interpret these results as the formation of superconducting spin gaps, consistent with previous reports for LSCO.

\section{Experimental Details}

High-quality single crystals of $La_{2-x}(Sr,Ba)_{x}CuO_{4}$ were grown by floating zone image furnace techniques using a four-mirror optical furnace. The growths followed the protocols already reported for the non-superconducting samples\cite{Fujita_PRB_2004, Dabkowska_Springer_2010,Ellis_PRB_2010}. 

LBCO samples at low doping, $x$ $\leq$ 0.05, such that they possess non-superconducting ground states, display orthorhombic crystal structures with space group {\em Bmab}\cite{Boni_PRB_1988,Kastner_RevModPhys_1998} at all temperatures measured in these experiments. At higher doping, $x$ $>$ 0.05, such that both LBCO and LSCO samples possess superconducting ground states, both orthorhombic and tetragonal crystal structures are observed over the temperature ranges measured\cite{Keimer_PRB_1992_2,Zhao_PRB_2007}. Despite this complexity in the structure of the materials studied, the distinction between the a and b lattice parameters within the orthorhombic structures is small, and in light of the relatively low {\bf Q} resolution of our measurements, we choose to approximate all of these crystal structures by the high temperature $I4/mmm$ tetragonal structure that is displayed by the parent compound, $La_{2}CuO_{4}$. We will therefore adopt the tetragonal notation for all our samples at all temperatures measured\cite{Katano_PhysicaC_1993,Lee_JPhysSocJpn_2000} in this study. All crystal structures within these families are layered which gives rise to quasi-two dimensional magnetism over most of the phase diagram. Consequently, magnetic zone centers are centered around equivalent ($\frac{1}{2},\frac{1}{2},L$) tetragonal wave-vectors, and appear extended along $L$. We will refer to these lines in reciprocal space as two dimensional magnetic zone centers (2DMZCs), and much of our focus in this paper will be on these features within reciprocal space.

Neutron scattering measurements were performed using the ARCS and SEQUOIA time-of-flight chopper spectrometers, which are both located at the Spallation Neutron Source at Oak Ridge National Laboratory\cite{Abernathy_RevSciInst_2012,Granroth_JPhysConfSer_2010}. Both are direct geometry chopper instruments and use the same ambient temperature moderator for their incident neutrons\cite{Haines_NIMR_2014}. The single crystal samples were mounted in closed cycle refrigerators allowing measurements to probe the approximate temperature range from 5 to 300 K with a temperature stability of $\sim$ 0.1 K. All measurements were performed with single crystal samples aligned such that their {\em HHL} scattering plane was horizontal. We employed $E_{i}$ = 60 meV incident energy neutrons for all measurements shown and employed single crystal sample rotation about a vertical axis. By coupling this single crystal sample rotation experimental protocol with the large, two dimensional detector arrays of ARCS and SEQUOIA, we obtained comprehensive four-dimensional master data sets in each experiment (3 {\bf Q} and 1 energy dimensions), which we can project into different scattering planes by appropriate integrations of the data. 

SEQUOIA was used to measure the $x$ = 0 and 0.05 LBCO samples. In these measurements, we employed SEQUOIA's 700 meV high flux chopper to select the incident neutron energy, 60 meV, resulting in an energy resolution at the elastic position of $\sim$ 1 meV. Measurements swept out 141 degrees of single crystal sample rotation, collected in 1 degree steps. Measurements at ARCS were performed on the LBCO $x$ = 0.035 and 0.095 and both LSCO samples. Here we employed ARCS' 100 meV chopper\cite{Stone_RevSciInst_2014} to select $E_{i}$ = 60 meV, and again the resulting energy resolution was $\sim$ 1 meV at the elastic position. These measurements swept out 140 degrees of single crystal sample rotation in one degree steps. All data reduction and analysis for this work were carried out using Mantid\cite{Mantid} and Horace\cite{Horace}, as appropriate.

\section{Contour Maps of the Scattered Neutron Intensity}

\begin{figure}
\centering 
\includegraphics[scale=0.1]{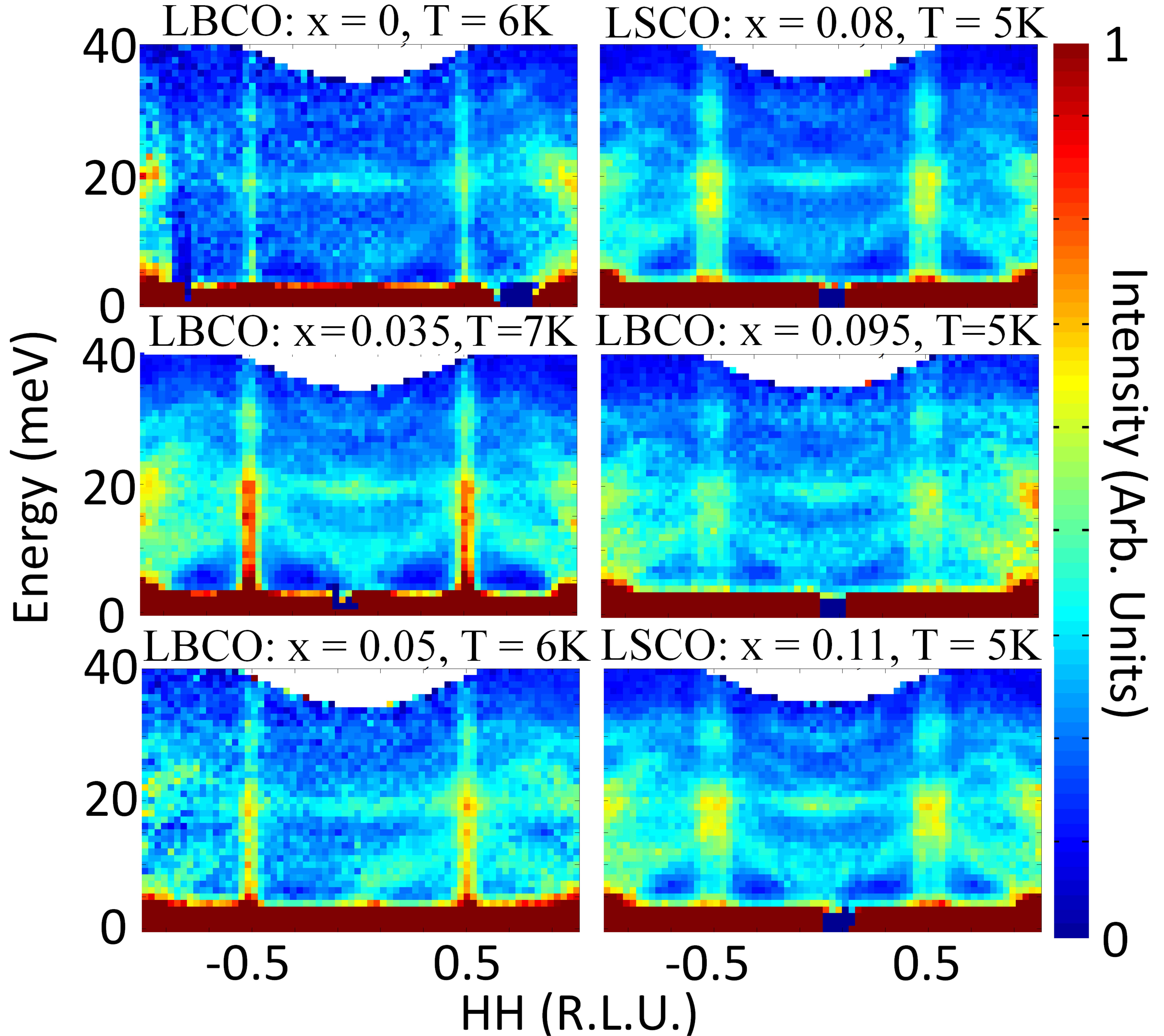} 
\caption{Energy vs. $HH$ maps for all samples measured, as labeled. The data shown employs the subtraction of an empty can data set\cite{Wagman_PRB_2015,Pintschovius_Arxiv_2014}, integration from -0.1 to 0.1 in $<H\bar{H}>$ and -4 to 4 in $<L>$. The vertical rod shaped features, emanating from ($\frac{1}{2}$,$\frac{1}{2}$) positions are the dispersive magnetic excitations. All data have been normalized to be on the same absolute intensity scale as described in the text.}
\label{E vs HH}
\end{figure}

Our time-of-flight neutron data sets span all four dimensions of energy-reciprocal space. As a result, in order to view projections of the scattering in different scattering planes, we must integrate about out-of-plane directions, as appropriate. Scattering planes, or so-called slices, are obtained by integrating the master data set about two out-of-plane directions. Constant-energy or constant-{\bf Q} cuts are obtained by integration of the master data set about three directions\cite{Wagman_PRB_2015}.

We first present energy vs. $HH$ maps of the scattering for all the single crystals measured at base cryostat temperature, which are between 5 and 7 K. These maps are obtained by integrating from -0.1 to 0.1 in $H\bar{H}$ and from -4 to 4 in $L$, and are presented in Fig. \ref{E vs HH} for all of our LBCO and LSCO samples, as labeled. We have also normalized each data set to the same absolute, but otherwise arbitrary, intensity scale by using a combination of normalization to incoherent elastic scattering and/or low energy acoustic phonon scattering at 6 meV, near the (0 0 16) Bragg peak\cite{Xu_RevSciInst_2013}.

From Fig. \ref{E vs HH} we see several common features for all the samples. The most salient common feature is the highly dispersive rod-shaped inelastic scattering that emanates from both of {\bf Q} = ($\pm\frac{1}{2},\pm\frac{1}{2},L$). These rods of inelastic scattering are the highly dispersive spin excitations. One notes a small drop off in this magnetic inelastic intensity with increased doping, although the LBCO $x$ = 0 magnetic scattering appears weak due the effects of experimental resolution and signal integration. Nonetheless this is a relatively weak effect and the overall magnetic spectral weight at energies less than $\sim$ 40 meV is not significantly diminished for doping levels out to $x$ $\sim$ 0.11. In addition, an increase in the breadth of the magnetic scattering along {\bf Q} is observed, which is consistent with a linear doping dependence of the incommensurate splitting of the magnetic excitations. Such a doping dependence is known to describe the incommensuration of the 2DMZCs\cite{Lee_RevModPhys_2006}. It should be noted that the inelastic magnetic scattering is understood to exhibit an hour-glass shaped dispersion\cite{Matsuda_PRB_2013,Matsuda_PRL_2008}. However, our relatively low {\bf Q} resolution measurement is not sensitive to such hour-glass features and the magnetic scattering appears instead as dispersive rods emanating from the 2DMZCs. The incommensurate nature of the inelastic scattering is pronounced and obvious in Fig. \ref{E vs HH} for all of the samples with superconducting ground states, which are those with $x$ $>$ 0.05. Several clear phonon branches can also be seen within this field of view. These are the quasi-2D phonons common to all of these materials, as previously discussed\cite{Wagman_PRB_2015}. As we are employing a rather large integration in $L$ ($\pm$ 4), we expect that three dimensional features will be averaged out by such an integration, while 2D features that are dispersionless along $L$, will present more clearly in such a plot.

\begin{figure}
\centering 
\includegraphics[scale=0.16]{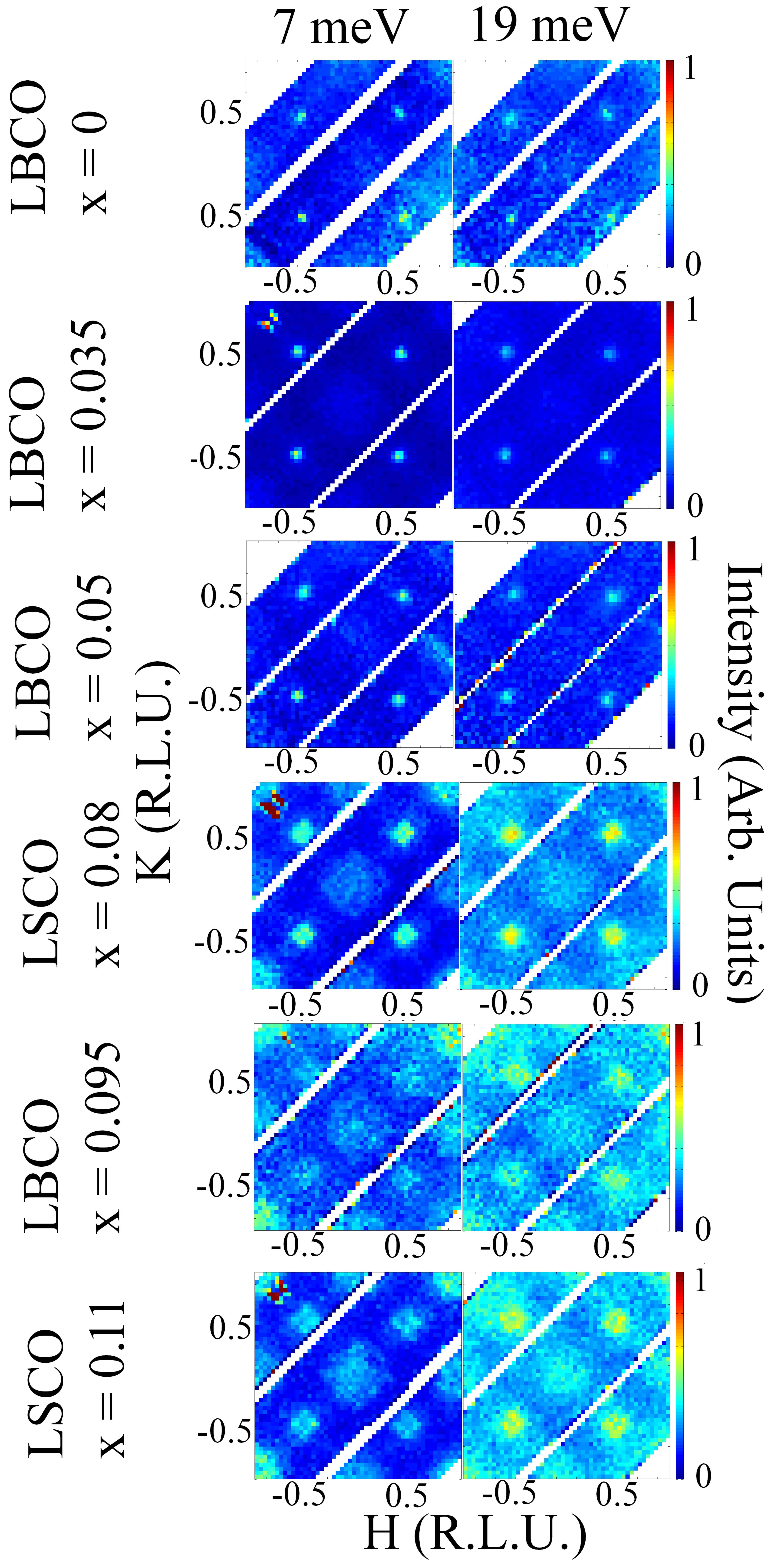} 
\caption{Maps of the scattering in the $HK$ plane for all samples measured, as labeled. The data shown employs integration from -4 to 4 in $<L>$ and $\pm$ 1 meV in energy, as labeled. Data have been normalized separately, as described in the text.}
\label{HK}
\end{figure}

\begin{figure}
\centering 
\includegraphics[scale=0.13]{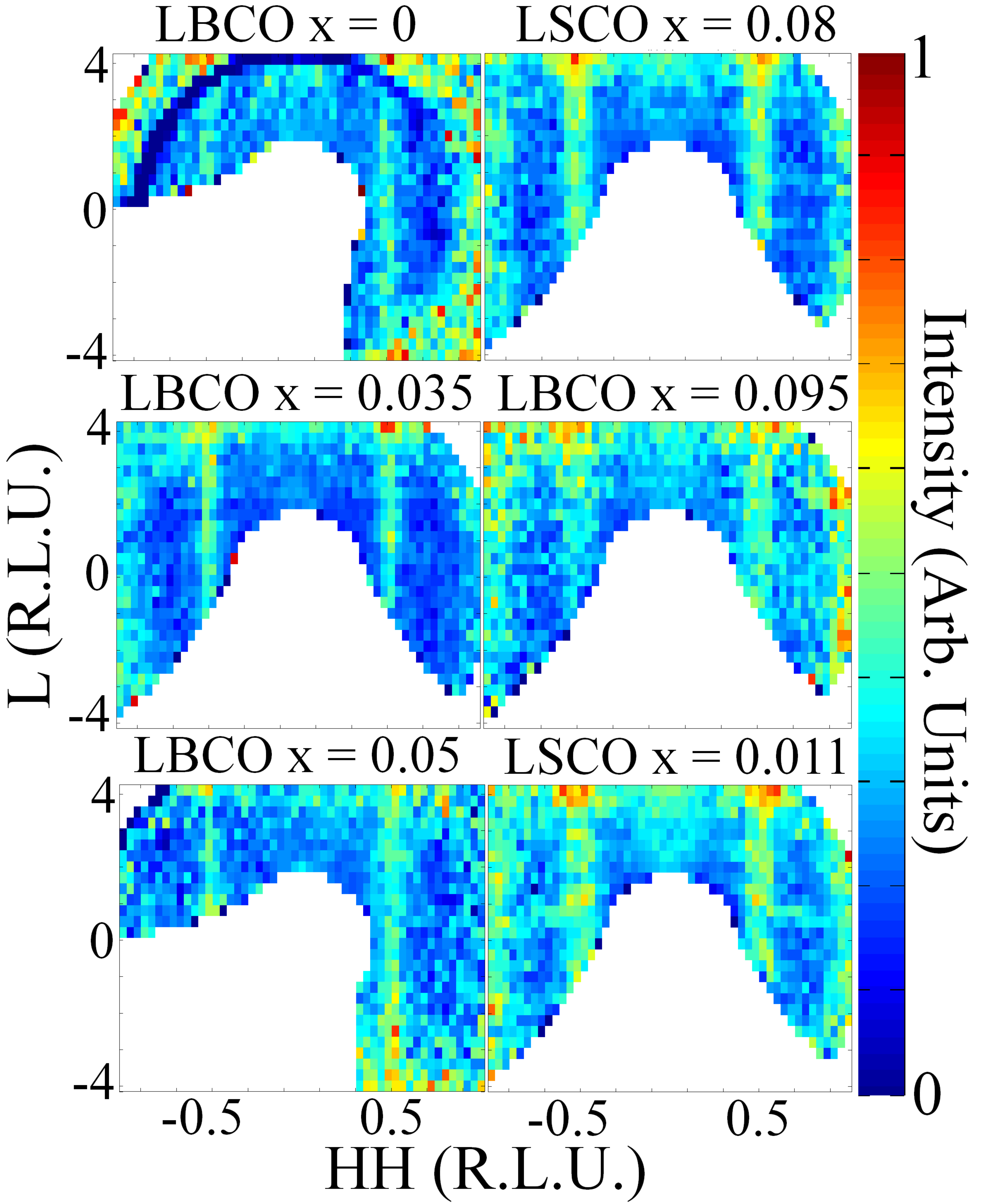} 
\caption{Maps of the scattering in the $HHL$ for all samples measured, as labeled. The data shown employs integration from -0.1 to 0.1 in $<H\bar{H}>$ and $\pm$ 1 meV about 19 meV. Data have been normalized to the same absolute, arbitrary scale.}
\label{HHL}
\end{figure}

\begin{figure}
	\centering 
	\includegraphics[scale=0.27]{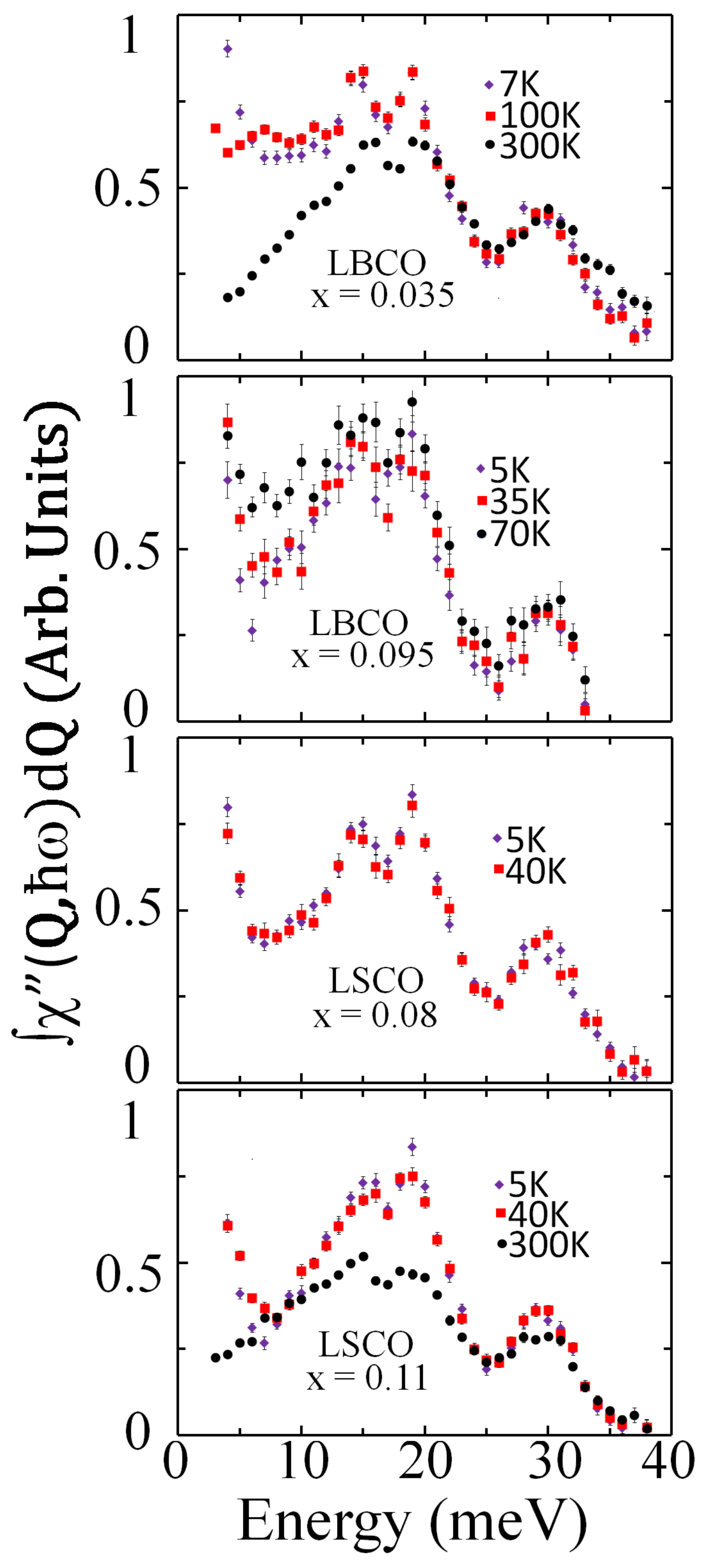} 
	\caption{Constant-energy cuts along the ($-\frac{1}{2},-\frac{1}{2}$) 2DZMC plotted at all measured temperatures for all samples. All data shown were integrated from -4 to 4 in $L$, -0.1 to 0.1 in $H\bar{H}$ and -0.6 to -0.4 in $HH$. The data has been normalized to the same absolute intensity scale, corrected for the Bose factor and employs a subtraction of a {\bf Q} and energy independent background, as described in the text. Error bars represent one standard deviation.}
	\label{TDep}
\end{figure}

Common to all six maps in Fig. \ref{E vs HH} is the strong enhancement of the inelastic scattering seen at the crossings of the dispersive spin excitations with the relatively dispersionless optic phonons. This enhancement has been previously discussed for the LBCO $x$ = 0.035 sample\cite{Wagman_PRB_2015}. Here we see a remarkably consistent phenomenology as a function of doping, for systems with both superconducting and non-superconducting ground states, and for both LBCO and LSCO. The enhanced inelastic scattering increases in breadth along {\bf Q}, consistent with an increased incommensuration of the magnetic inelastic scattering emanating from 2DMZCs as a function of doping, as is expected. 

We now turn to constant energy slices of the $HK$ plane in Fig. \ref{HK}, again derived from our master 4 dimensional data set. To obtain this projection, we again integrate from -4 to 4 in $L$ but now integrate by $\pm$ 1 meV in energy. We have done this for all six data sets shown at two energies - 7 meV, an energy at which the scattering at the lowest $|{\bf Q}|$ 2DMZCs is almost entirely comprised of magnetic scattering, and 19 meV, the energy for which the optic phonons in the 214 cuprates are quasi-2D in nature and where the enhanced scattered intensity is maximal. Here, we do not normalize each data set to a single absolute, arbitrary intensity scale. Instead, we normalize each data set such that their respective intensity scales at 7 meV appear qualitatively similar, and we then employ the same normalization for the corresponding 19 meV data sets.

Consider first the left column of Fig. \ref{HK}. This shows the 7 meV data for all six samples measured. At this energy, there are no crossings of phonons with the spin excitations at the 2DMZCs. At the lowest $|{\bf Q}|$ 2DMZC we expect minimal contributions from phonon scattering such that the scattered intensity is magnetic in origin. The extent of the scattering within the $HK$ plane increases with doping, x, although it is most noticeable for $x$ $>$ 0.05. We also note that the ratio of the magnetic scattering around the 2DMZC to the nearby background scattering, which is comprised of phonon scattering, decreases as a function of $x$, albeit only slowly. Some decrease in the magnetic scattering with increased $x$ is expected, as magnetic moments are being removed from the samples. Such an effect should appear at least linearly with $x$\cite{Enoki_PRL_2013,Fujita_PRB_2002,Wakimoto_PRB_2001_n}. Nonetheless, this data, and those shown in Fig. \ref{E vs HH}, make it clear that significant dynamic magnetic spectral weight is present well into the $La_{2-x}(Sr,Ba)_{x}CuO_{4}$ phase diagram, and clearly coexists with superconductivity.

Turning to the $HK$ slices at 19 meV, shown in the right column of Fig. \ref{HK}, we see similar trends to those seen at 7 meV. We find that the extent of the scattering within the $HK$ plane increases with doping in much the same way as is observed at 7 meV, and the relative strength of the scattering at 19 meV compared with 7 meV appears to increase with $x$. 

Figure \ref{HHL} focuses on this 19 meV scattering by projecting our 4 dimensional master data set into the $HHL$ scattering plane. In this figure, we again normalize using an absolute, arbitrary intensity scale. We clearly see isotropic rods of scattering that extend along L for the 2DMZCs of the form ($\frac{1}{2}, \frac{1}{2}, L$). Such rods of scattering are indicative of the 2D nature of the enhancements seen in Fig. \ref{E vs HH}. We clearly identify the increasing extent of the rods of scattering in the $HH$ direction with $x$, and see that this occurs along the full rod of scattering along $L$. 

\section{Analysis and Discussion}

Taken together, Figs. 1-3 show consistent phenomenology across the underdoped region of the 214 cuprate phase diagram, out to almost $x$ = $\frac{1}{8}$. We now focus on a quantitative analysis of the energy dependence of the spectral weight emanating from the 2DMZCs and the resonant enhancement of this spectral weight coincident with crossings of the spin excitations and low-lying optic phonons, as previously reported for LBCO with $x$ = 0.035\cite{Wagman_PRB_2015}. We convert our measured S({\bf Q},$\hbar\omega$) to the imaginary part of the susceptibility, or $\chi\prime\prime({\bf Q},\hbar\omega)$ using similar protocols to those used for the LBCO $x$ = 0.035 analysis\cite{Wagman_PRB_2015}. The relationship between S({\bf Q},$\hbar\omega$) and $\chi\prime\prime({\bf Q},\hbar\omega)$ is given by: 

\begin{equation}
S({\bf Q},\omega,T) = [n(\hbar\omega)+1)]\times\chi\prime\prime({\bf Q},\omega,T)
\end{equation}

where

\begin{equation}
[n(\hbar\omega) + 1)] = \frac{1}{1-e^{-\frac{\hbar\omega}{k_{B}T}}}
\end{equation}

is commonly referred to as the Bose factor\cite{Squires_Text}. To compare the dynamic susceptibility appropriately, one must remove background contributions to the scattered intensity. We employ the same form of background subtraction as was previously used for LBCO, $x$ = 0.035\cite{Wagman_PRB_2015}. For each sample, we first employ an integration from -4 to 4 in L and -0.1 to 0.1 in $\bar{H}H$. From there, we further integrate in $HH$ from $\pm$ 0.2 to $\pm$ 0.4 and $\pm$ 0.6 to $\pm$ 0.8 in $HH$ to give us a measure of the background away from the 2DMZCs but bounded by the nearby acoustic phonon, as can be seen in Fig. \ref{E vs HH} for all of our data sets. Having accounted for the experimental background, we remove the Bose factor from our data and normalize our data sets to an absolute scale. We then quantitatively compare the energy dependence of the $\bf Q$-integrated (around the 2DMZC) $\chi\prime\prime({\bf Q},\omega,T)$ as a function of doping, $x$ in Figs. 4, 5 and 6. 

\begin{figure}
	\centering 
	\includegraphics[scale=0.28]{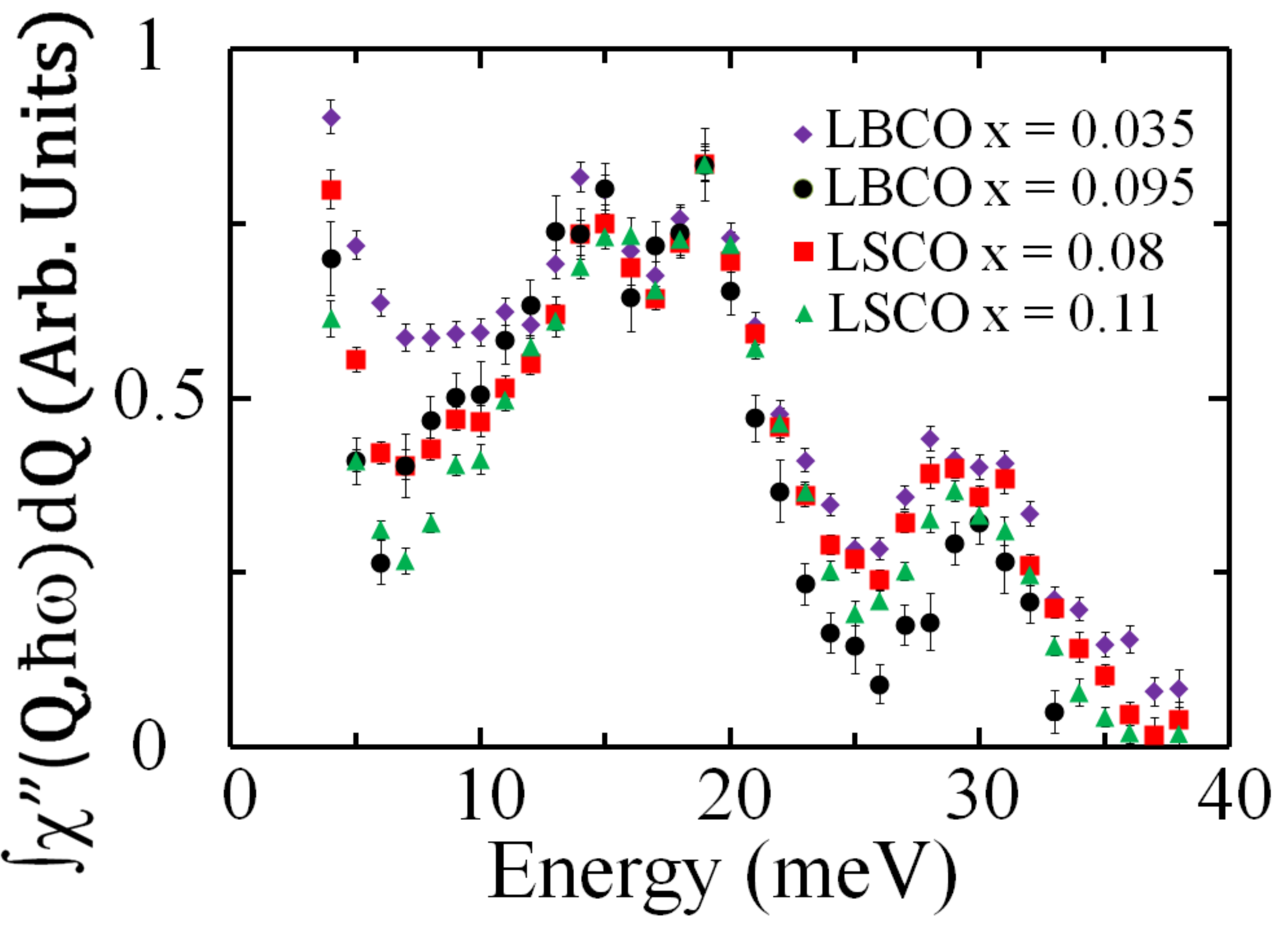} 
	\caption{Constant-energy cuts along the ($-\frac{1}{2},-\frac{1}{2}$) direction, as shown in Fig. \ref{TDep}, for the lowest temperature data sets collected on each sample. The data employ the same -4 to 4, -0.6 to -0.4 and -0.1 to 0.1 in $L$, $HH$ and $\bar{H}H$, respectively. Here, all data shown have normalized to the same arbitrary intensity scale. Error bars represent one standard deviation.}
	\label{DDep}
\end{figure}

We focus on the lowest $|{\bf Q}|$ 2DMZC {\bf Q} = ($-\frac{1}{2},-\frac{1}{2}$) position,  and employ a relatively wide integration in L, from -4 to 4, so as to effectively capture the quasi-2D scattering. We also compare data sets taken on ARCS only, as there are four such data sets that span the key range of the 214 cuprate phase diagram, and these allow us the most ``like-with-like" comparison of our data sets.

Figure \ref{TDep} shows the integrated dynamic susceptibility, $\chi\prime\prime({\bf Q},\hbar\omega)$, for all four samples measured on ARCS and at all temperatures investigated. These are all of our samples with superconducting ground states and one sample with a non-superconducting ground state (LBCO $x$ = 0.035). All of these data sets show very similar parametric behavior above $\sim$ 10 meV. We find that the effects of temperature do not significantly affect the scattering above 10 meV until the temperature reaches on the order of 300 K. At 300 K $\chi\prime\prime({\bf Q},\hbar\omega)$ is noticeably reduced especially below $\sim$ 15 meV. The bottom three panels of Fig. \ref{TDep} all show the integrated dynamic susceptibility $\chi\prime\prime({\bf Q},\hbar\omega)$ for underdoped LBCO and LSCO samples with superconducting ground states. In addition these plots all show data sets at T = 5 K, which is well below each sample's respective superconducting T$_{C}$, and at T = 35 K or 40 K, which are around 5 K above each sample's respective T$_{C}$. 

Figure 5 shows the integrated dynamic susceptibility, $\chi\prime\prime({\bf Q},\hbar\omega)$ at low temperatures for all four samples shown in Fig. \ref{TDep}, but now overlaid such that the similarities and differences between low temperature $\chi\prime\prime({\bf Q},\hbar\omega)$ as a function of doping, $x$, can be explicitly seen. Normalizing the $\chi\prime\prime({\bf Q},\hbar\omega)$ to agree at all dopings in the resonant enhancement energy regime, 15 - 20 meV, we see that the integrated dynamic susceptibility, $\chi\prime\prime({\bf Q},\hbar\omega)$ at low temperatures agree in detail remarkably well at all energies from $\sim$ 10 meV to 25 meV, for the LBCO and LSCO samples with superconducting ground states, $x$ = 0.08, 0.095 and 0.11. The LBCO sample with a non-superconducting ground state, $x$ = 0.035, agrees with the other samples very well above $\sim$ 12 meV, but shows enhanced magnetic spectral weight at energies below $\sim$ 12 meV. The overall phenomenology is clear; the integrated dynamic susceptibility, $\chi\prime\prime({\bf Q},\hbar\omega)$ at low temperatures is very similar for underdoped LBCO and LSCO at all doping levels measured, with the proviso that there is enhanced low energy ($<$ 12 meV) magnetic spectral weight for the non-superconducting $x$ = 0.035 sample.

The quantitative agreement between the integrated dynamic susceptibility, $\chi\prime\prime({\bf Q},\hbar\omega)$ at low temperatures and below $\sim$ 35 meV across over such a large range of doping in both LBCO and LSCO is remarkable. Combined with the earlier observation from Figs. 1-3 that the breadth in {\bf Q} of the enhancements track with the incommensuration about the 2DMZC, while staying centred on the energies of the low lying optic phonons, we are led to an interpretation of the enhancement that depends on both the spin and phonon degrees of freedom. Such an effect would likely involve a hybridization of quasi-2D spin degrees of freedom with optic phonons, as opposed to a solely magnetic origin.

\begin{figure}
	\centering 
	\includegraphics[scale=0.42]{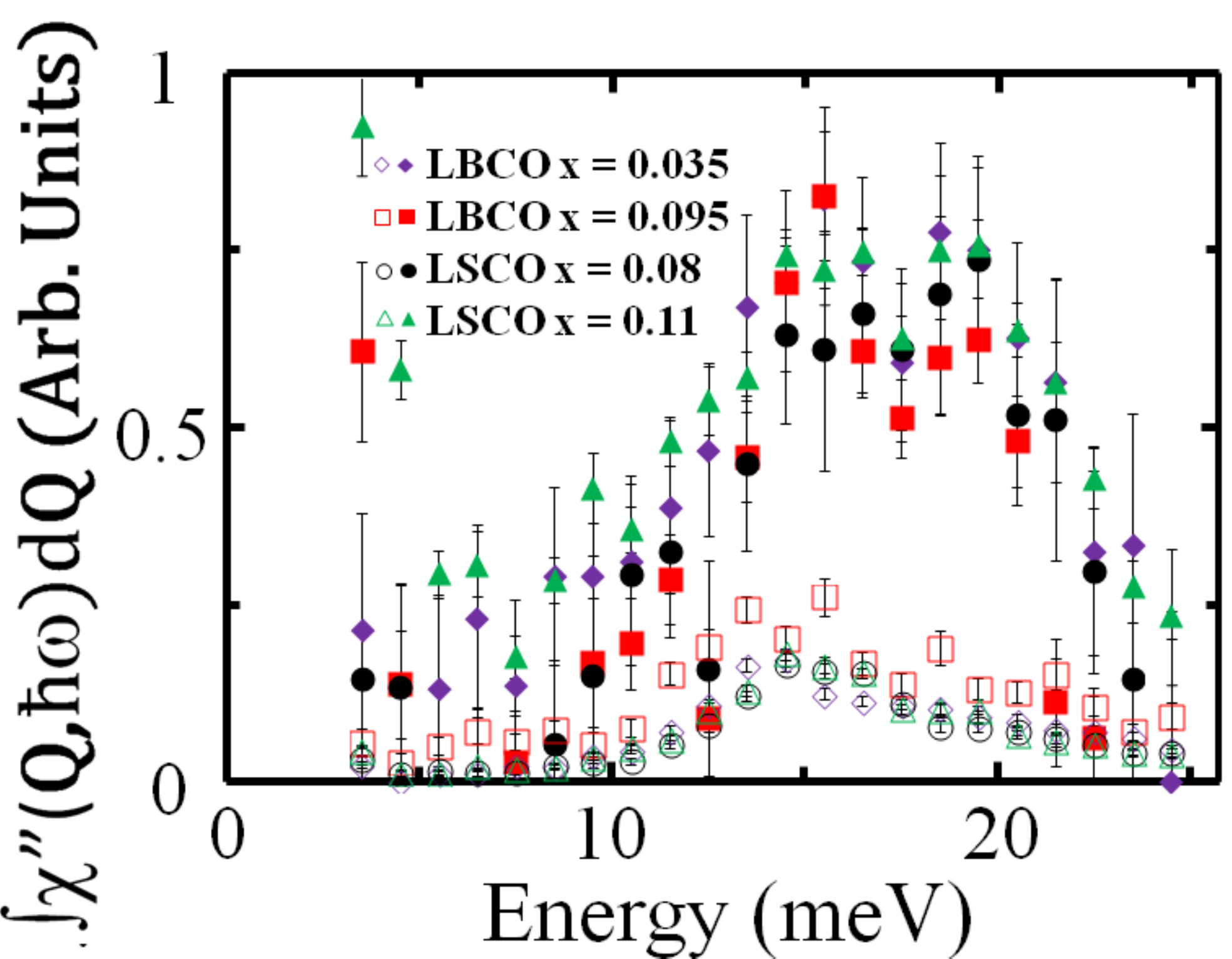} 
	\caption{$|Q|^{2}$ normalized integrated $\chi\prime\prime(\hbar\omega,{\bf Q})$ for all ARCS data sets, as described in the text. A narrow $L$ integration of -0.5 to 0.5 and $\pm$0.1 in both $HH$ and $\bar{H}H$ about the ($-\frac{1}{2},-\frac{1}{2},0$) and ($-\frac{5}{2},-\frac{5}{2},0$) 2DMZCs is employed for all samples measured. Closed symbol data sets correspond to data from {\bf Q} = ($-\frac{1}{2},-\frac{1}{2}$), while open symbol data sets correspond to data from {\bf Q} = ($-\frac{5}{2},-\frac{5}{2}$). Error bars represent one standard deviation.}
\end{figure}

\begin{figure}
	\centering 
	\includegraphics[scale=0.17]{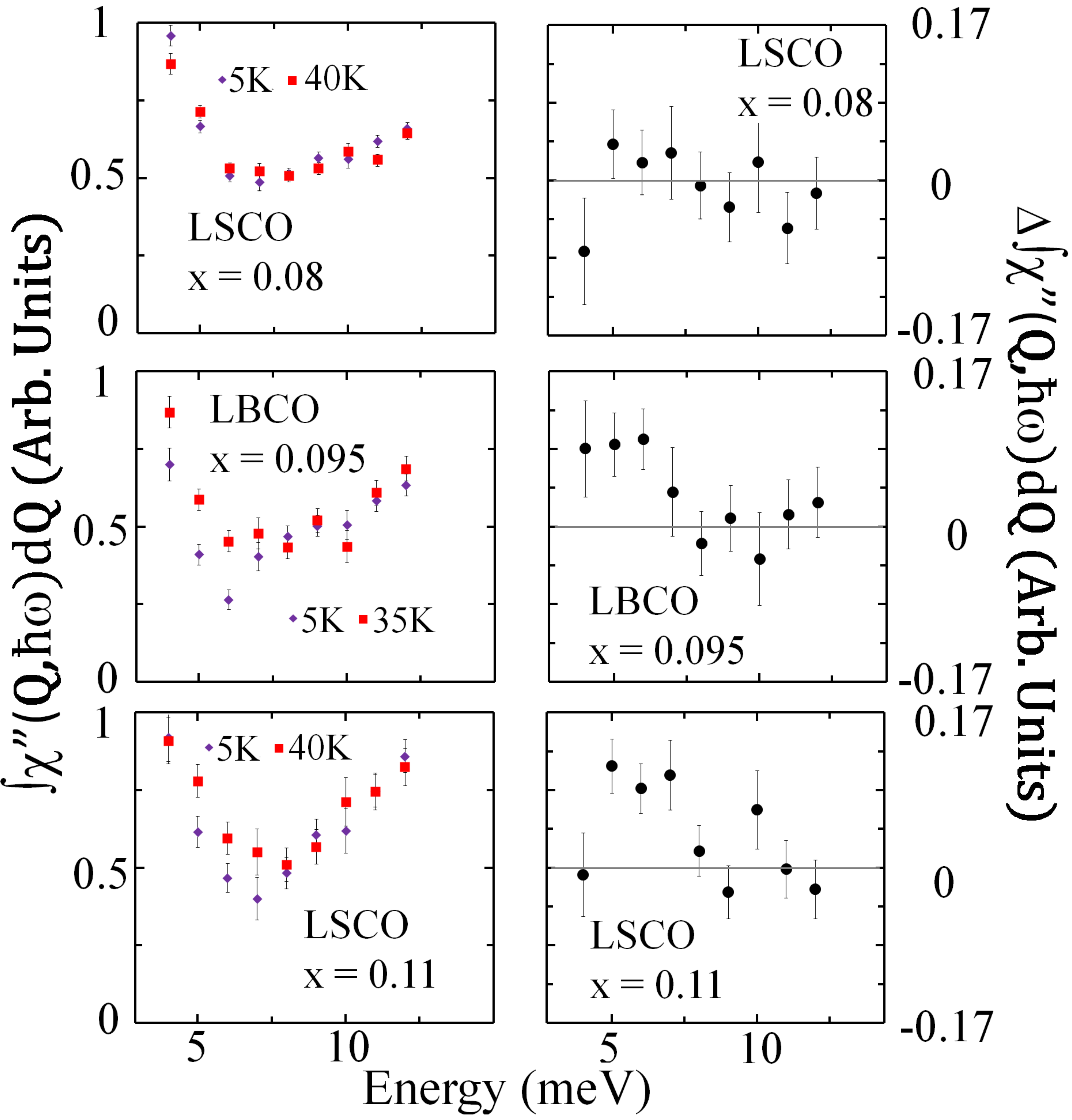} 
	\caption{Left Column: Integrated $\chi\prime\prime({\bf Q},\hbar\omega)$ for the three samples with superconducting ground states. These data have been integrated from -4 to 4 in $L$, from -0.1 to 0.1 in $H\bar{H}$ and from -0.6 to -0.4 in $HH$. Only a {\bf Q} and energy independent background has been subtracted from the data set. Right Column: Difference plots between the high temperature (35 K or 40 K) and the low temperature (5 K) data sets shown in the left column of this figure. Data sets from the same sample (in the right or left column) employ the same arbitrary intensity scale. Error bars represent one standard deviation.}
	\label{SG}
\end{figure}

As was done previously for LBCO x = 0.035\cite{Wagman_PRB_2015}, we can compare the strength and form of $\chi\prime\prime({\bf Q},\hbar\omega)$ as a function of {\bf Q} at 2DMZCs for which the nuclear structure factor is identical (within the $I4/mmm$ space group). The structure factors are identical at wave vectors of the form ($\frac{H}{2}, \frac{H}{2}, 0$) and in Fig. 6, we compare $\chi\prime\prime({\bf Q},\hbar\omega)$ integrated around the ($-\frac{1}{2}, -\frac{1}{2}, 0$) and ($-\frac{5}{2}, -\frac{5}{2}, 0$) wave-vectors. For this comparison we employ a relatively narrow integration in $L$ about $L$ = 0, from -0.5 to 0.5. We observe the same large enhancements to $\chi\prime\prime({\bf Q},\hbar\omega)$ near 15 meV and 19 meV around ($-\frac{1}{2}, -\frac{1}{2}, 0$) as were seen in Figs. 4 and 5. Were this enhancement due solely to phonons, it would scale as $|{\bf Q}|^{2}$. We have scaled the measured $\chi\prime\prime({\bf Q},\hbar\omega)$ by $|{\bf Q}|^{2}$ in Fig. 6, and clearly the $|{\bf Q}|^{2}$ scaled $\chi\prime\prime({\bf Q},\hbar\omega)$ is much stronger near ($-\frac{1}{2}, -\frac{1}{2}, 0$) than near ($-\frac{5}{2}, -\frac{5}{2}, 0$). This eliminates the possibility that the enhancement is due to phonons alone, or due to a simple superposition of phonons and spin excitations whose spectral weight monotonically decreases with energy.  Fig. 6 shows that such a conclusion follows for all concentrations of LBCO and LSCO studied.

Finally, we address the issue of whether or not a spin gap, a suppression in the magnetic spectral weight at low energies, occurs in underdoped LBCO and LSCO on reducing temperature and entering the superconducting state. As can be seen in Fig. \ref{TDep}, the presence of a spin gap will be a subtle effect. As the magnetic scattering is quasi-2D, we perform a similar analysis to that which produced Figs. 4 and 5, using a large integration in $L$ from -4 to 4 to better capture the quasi-2D magnetic scattering. The resulting integrated dynamic susceptibility, $\chi\prime\prime({\bf Q},\hbar\omega)$ is shown in Fig. 7 for our three samples with superconducting ground states, for energies below $\sim$ 10 meV, and for temperatures just above (35 K or 40 K) and well below (5 K), each sample's superconducting T$_{C}$. Data in the left hand column of Fig. 7 shows the integrated dynamic susceptibility, $\chi\prime\prime({\bf Q},\hbar\omega)$ for the three crystals, while that in the right hand column of Fig. 7 shows the corresponding difference in integrated dynamic susceptibility between the superconducting (T = 5 K) and normal states (T = 35 K or 40 K). 

In this context, a spin gap is identified as excess integrated dynamic susceptibility, $\chi\prime\prime({\bf Q},\hbar\omega)$, occurring at low energies in the higher temperature normal state, as compared to the lower temperature superconducting state. While the effect of the spin gap is subtle, our data is consistent with a spin gap of $\sim$ 8 meV for $x$ = 0.11, falling to $\sim$ 2 meV or lower for $x$ = 0.08. Presumably, the spin gap energy should fall to zero at the low $x$ onset of superconductivity in these families, which is $x$ = 0.05. We note that the superconducting spin gap we observe in LBCO $x$ = 0.095 is similar but $\sim$ 1 meV lower than that displayed in LSCO $x$ = 0.11. Our results show consistency between the LBCO and LSCO families, as expected as their physical properties are so similar. The observation of a spin gap in LBCO resolves a long-standing puzzle that LBCO had not previously shown a spin gap, while LSCO had\cite{Tranquada_Text2}. For LSCO $x$ = 0.11, the spin gap energy scale appears to be consistent with previous reports, with a gap energy around 8 meV\cite{Christensen_PRL_2004,Lake_Nature_1999}. While there does not appear to be a gap in the presented LSCO $x$ = 0.08 data, we believe this to be a result of the spin gap energy being below 2 meV.

\section{Conclusions}

We have carried out comprehensive inelastic neutron scattering measurements using single crystal sample rotation and time of flight techniques on samples of the underdoped 214 cuprate superconductors, LBCO and LSCO, for doping levels between $x$ = 0 and $x$ = 0.11. All of these samples show a resonant enhancement of the inelastic spectral weight at 2DMZCs and at energies which correspond to crossings of the highly dispersive spin excitations with weakly dispersive optic phonons. These results are quantitatively similar to those previously reported for non-superconducting LBCO with $x$ = 0.035, but which are now extended well into the superconducting part of the LBCO and LSCO phase diagrams. This enhancement is therefore a generic property of these families of quasi two dimensional, single layer copper oxides. 

While it is possible that the enhanced spectral weight as a function of energy at 2DMZCs is a purely magnetic effect, as was postulated earlier for LSCO with $x$ = 0.085 and 0.016\cite{Vignolle_NaturePhys_2007,Lipscombe_PRL_2007}, its occurence at the confluence in {\bf Q} and energy of dispersive spin excitations with optic phonons, and its doping independence, at least for $x$ $<$ 0.12, makes a hybridized spin-phonon resonance much more plausible. Furthermore, the eigenvector of the $\sim$ 19 meV optic phonon for which this enhancement is largest is known to be a quasi-two dimensional oxygen breathing mode, with ionic displacements primarily within the CuO$_2$ planes, as reported previously for LBCO with $x$ = 0.035. Such an eigenvector flexes the Cu-O bonds most responsible for strong antiferromagnetic superexchange, and such a phonon would be expected to couple strongly to magnetism in LBCO and LSCO.

If the requirements for this resonant enhancement are indeed dispersive spin excitations and quasi-two dimensional optic phonons capable of coupling strongly to the spin degrees of freedom, then we do expect this same phenomenology to persist across the copper oxide phase diagram, to samples with superconducting ground states, as we are reporting. This opens up the very real possibility that such an enhancement should exist in other families of high T$_C$ oxides, and the more speculative possibility that such a hybridized spin-phonon excitation plays a role in superconducting pairing.

We further show that the quantitative form of the low temperature, integrated dynamic susceptibility, $\chi\prime\prime({\bf Q},\hbar\omega)$ at the 2DMZC is very similar as a function of doping, at least out to $x$ = 0.11 in both LBCO and LSCO. The main changes that occur on doping is the suppression of magnetic spectral weight for energies less than $\sim$ 12 meV at low, non-superconducting dopings compared with higher, superconducting dopings and the development of a superconducting spin gap for $x$ $>$ 0.05 for both LBCO and LSCO.

\begin{acknowledgments}
We would like to acknowledge useful conversations had with N. Christensen, E. Taylor, J. P. Carbotte, T. Timusk, J. Tranquada, I. Zaliznyak and D. Fobes. We would also like to acknowledge T. E. Sherline and L. DeBeer Schmidt for technical assistance with the measurements on SEQUOIA, J. Niedziela and D. Maharaj for technical assistance with the ARCS measurements and E. McNeice for assistance with sample growth. Research using ORNL's Spallation Neutron Source was sponsored by the Scientific User Facilities Division, Office of Basic Energy Sciences, U.S. Department of Energy. Work at McMaster was funded by NSERC of Canada.
\end{acknowledgments}

\bibliographystyle{unsrt}
\bibliography{cuprate}

\end{document}